\documentstyle[12pt,epsf,cite]{article}

\makeatletter
\@addtoreset{equation}{section}
\makeatother

\def\ben{\begin{equation}}
\def\een{\end{equation}}

\let\a=\alpha    
    
  \let\n=\nu

\let\C=\Chi

\def\nn{\nonumber} \def\bd{\begin{document}} \def\ed{\end{document}}
\def\ds{\documentstyle} \let\fr=\frac \let\bl=\bigl \let\br=\bigr
\let\Br=\Bigr \let\Bl=\Bigl
\let\bm=\bibitem
\let\na=\nabla
\let\pa=\partial \let\ov=\overline
\newcommand{\be}{\begin{equation}}
\newcommand{\ee}{\end{equation}}
\def\ba{\begin{array}}
\def\ea{\end{array}}
\def\ft#1#2{{\textstyle{{\scriptstyle #1}\over {\scriptstyle #2}}}}
\def\fft#1#2{{#1 \over #2}}
\def\del{\partial}
\def\vp{\varphi}
\def\sst#1{{\scriptscriptstyle #1}}
\def\oneone{\rlap 1\mkern4mu{\rm l}}
\def\td{\tilde}
\def\wtd{\widetilde}
\def\ie{\rm i.e.\ }
\def\dalemb#1#2{{\vbox{\hrule height .#2pt
        \hbox{\vrule width.#2pt height#1pt \kern#1pt
                \vrule width.#2pt}
        \hrule height.#2pt}}}
\def\square{\mathord{\dalemb{6.8}{7}\hbox{\hskip1pt}}}
\newcommand{\ho}[1]{$\, ^{#1}$}
\newcommand{\hoch}[1]{$\, ^{#1}$}
\newcommand{\bea}{\begin{eqnarray}}
\newcommand{\eea}{\end{eqnarray}}
\newcommand{\ra}{\rightarrow}
\newcommand{\lra}{\longrightarrow}
\newcommand{\Lra}{\Leftrightarrow}
\newcommand{\ap}{\alpha^\prime}
\newcommand{\bp}{\tilde \beta^\prime}
\newcommand{\tr}{{\rm tr} }
\newcommand{\Tr}{{\rm Tr} }
\def\0{{\sst{(0)}}}
\def\1{{\sst{(1)}}}
\def\2{{\sst{(2)}}}
\def\3{{\sst{(3)}}}
\def\4{{\sst{(4)}}}
\def\5{{\sst{(5)}}}
\def\6{{\sst{(6)}}}
\def\7{{\sst{(7)}}}
\def\8{{\sst{(8)}}}
\def\n{{\sst{(n)}}}
\def\cA{{{\cal A}}}
\def\cB{{{\cal B}}}
\def\cF{{{\cal F}}}
\def\tV{\widetilde V}
\def\tW{\widetilde W}
\def\tH{\widetilde H}
\def\tE{\widetilde E}
\def\tF{\widetilde F}
\def\tA{\widetilde A}
\def\im{{{\rm i}}}
\def\tY{{{\wtd Y}}}
\def\ep{{\epsilon}}
\def\vep{{\varepsilon}}
\def\R{\rlap{\rm I}\mkern3mu{\rm R}}
\def\bD{{{\bar D}}}

\def\R{\rlap{\rm I}\mkern3mu{\rm R}}
\def\bD{{{\bar D}}}
\def\R{{{\Bbb R}}}
\def\C{{{\Bbb C}}}
\def\H{{{\Bbb H}}}
\def\CP{{{\Bbb C}{\Bbb P}}}
\def\RP{{{\Bbb R}{\Bbb P}}}
\def\Z{{{\Bbb Z}}}
\def\bA{{{\Bbb A}}}
\def\bB{{{\Bbb B}}}
\def\bC{{{\Bbb C}}}
\def\bD{{{\Bbb D}}}
\def\bE{{{\Bbb E}}}
\def\bZ{{{\Bbb Z}}}
\def\Re{{{\frak{Re}}}}
\def\Im{{{\frak{Im}}}}
\def\cosec{{\,\hbox{cosec}\,}}
\def\Gm{{\Gamma_{\!\! -}}}
\def\Gp{{\Gamma_{\!\! +}}}
\def\stan{{standard }}
\def\nonstan{{supernumerary }}

\newcommand{\tamphys}{\it Center for Theoretical Physics,
Texas A\&M University, College Station, TX 77843}

\newcommand{\upenn}{\it Department of Physics and Astronomy,\\ University
of Pennsylvania, Philadelphia, PA 19104}

\newcommand{\brussels}{\it Physique Th\'eorique et Math\'ematique,
Universit\'e Libre de Bruxelles,\\ Campus Plaine C.P. 231, B-1050
Bruxelles, Belgium}

\newcommand{\auth}{H. L\"u\hoch{\dagger\star12},
C.N. Pope\hoch{\dagger1} and J.F.
V\'azquez-Poritz\hoch{\ddagger3}}

\begin{document}
\begin{flushright}

MIFP-03-15\ \ \ \ \
USTC-ICTS-03-11\ \ \ \ \
ULB-TH/03-24\\
July  2003\ \ \ \ \ {\bf hep-th/0307001}\\
\end{flushright}

\vspace{5pt}

\begin{center}

{\large {\bf From AdS Black Holes to Supersymmetric Flux-branes}}

\vspace{10pt}
\auth

\vspace{10pt}
{\hoch{\dagger}\it George P. and Cynthia W. Mitchell Institute for
Fundamental Physics,\\ Texas A\& M University, College Station, TX
77843-4242, USA}

\vspace{5pt}
{\hoch{\star}\it Interdisciplinary Center for Theoretical Study\\
University of Science and Technology of China, Hefei, Anhui 230026}

\vspace{5pt}
{\hoch{\ddagger}\brussels}

\vspace{40pt}

\underline{ABSTRACT}
\end{center}

     We show that AdS black hole solutions admit an analytical
continuation to become magnetic flux-branes.  Although a BPS AdS
black hole generally has a naked singularity, the BPS flux-brane
can be regular everywhere with an appropriate choice of
$U(1)$-charges. This flux-brane interpolates from AdS$_{D-2}
\times H^2$ at small distance to an asymptotic AdS$_D$-type metric
with an AdS$_{D-2}\times S^1$ boundary.  We also obtain a smooth
cosmological solution of de Sitter Einstein-Maxwell gravity which
flows from dS$_2\times S^{D-2}$ in the infinite past to a dS$_D$-type
metric, with an $S^{D-2}\times S^1$ boundary, in the infinite future.

{\vfill\leftline{}\vfill \vskip 10pt \footnoterule {\footnotesize
\hoch{1} Research supported in part by DOE grant
DE-FG03-95ER40917.

\hoch{2} Research supported in part by the USTC interdisciplinary
center for theoretic study through \phantom{grants} grants from the
Chinese Academy of Sciences and the Chinese NSF.

{\footnotesize \hoch{3} Research supported in full by the Francqui
Foundation (Belgium), the Actions de Recherche \phantom{of the}
Concert{\'e}es of the Direction de la Recherche Scientifique -
Communaut\'e Francaise de Belgique, \phantom{of the} IISN-Belgium
(convention 4.4505.86) and by a ``Pole d'Attraction
Interuniversitaire.''  }} \vskip  -12pt } \pagebreak
\setcounter{page}{1}

\newpage

\section{Introduction}

        According to the AdS/CFT correspondence \cite{malda,gkp,wit},
solutions in gauged supergravities can provide dual pictures of
certain quantum field theories.  Supersymmetric and regular solutions
are of particular interest, since then the supergravity approximation
can be trusted.  However, in both non-gauged and gauged
supergravities, solutions with these properties are rare.  In the
latter case, one expects that the geometry of the solution should be
asymptotic to AdS spacetime at large distance.  Thus, the regularity
of the solution depends on its small-distance behavior.  The possible
short-distance regular geometries are clearly limited.  One
possibility is AdS$_D$, with a different cosmological constant from
the large-distance AdS$_D$.  Such a solution, which is called a domain
wall, is typically supported by a scalar potential with two fixed
points.  This type of potential is quite rare.  An example in $D=5$
gauged supergravity was given in \cite{fgpw}.  A classification of
$D=4$ domain walls was given in \cite{cgr,cs}.  Another possibility is
that the small-distance geometry is AdS$_{D-n}\times S^n$ or
AdS$_{D-n}\times H^n$.  Supersymmetric solutions with $H^2$ have been
constructed in
\cite{romans,sabra3,sabra4,nunez,gauntlett4,sabra1,sabra2,klemm,%
sorin1,sorin2}, and recent examples with $S^2$ were given in
\cite{sorin1,sorin2}\footnote{Solutions with $n>2$, for $H^n$ mostly,
were constructed in
\cite{gauntlett1,gauntlett2,nunez2,gauntlett3}.}. These solutions are
regular everywhere, since they interpolate from AdS$_{D-2}\times S^2$
or AdS$_{D-2}\times H^2$ at small distance to an asymptotic AdS$_D$
form.

   These solutions are supported by a set of $U(1)$ gauge fields
carrying magnetic charges.  Their electric duals, namely the AdS black
holes, were constructed earlier
\cite{bh1,bh2,bh3,bcs,bh5,tenauthors,lm}.  By contrast, the electric
solutions are typically singular.  In this paper, we observe that
there exists an analytical continuation in which the electric BPS AdS
black hole becomes a supersymmetric magnetic flux-brane. Unlike the
AdS black hole, the flux-brane is regular everywhere, interpolating
between AdS$_{D-2}\times H^2$ at small distance to AdS$_D$ with a
boundary of AdS$_{D-1}\times S^1$ in the large-distance asymptotic
region.

    Thus, we see that the same AdS$_{D-2}\times H^2$ at small distance
can interpolate to two different AdS$_D$-type spacetimes at large
distance. For the magnetic $(D-3)$-branes constructed in
\cite{romans,sabra3,sabra4,nunez,gauntlett4,sabra1,sabra2,%
klemm,sorin1,sorin2}, the boundary of the AdS$_D$-type spacetime
is M$_{D-3}\times H^2$, whilst for the flux-branes presently
constructed, the boundary is AdS$_{D-1}\times S^1$.  To understand
why these two possibilities exist, let us write down the metric of
AdS$_{D-2}\times H^2$, given by
\bea ds_D^2 = d\rho^2 + e^{2\gamma\, \rho} dx^2 + dz^2 +
e^{2\wtd\gamma\, z}\, (-dt^2 + dx^i\, dx^i)\,. \eea
There are two directions which can act as a ``radial''-type
coordinate, namely $\rho$ and $z$, providing two different possible
interpolation coordinates. Previous works have established that
there can be an interpolation between the above AdS$_{D-2}\times
H^2$ metric at the horizon and an AdS$_D$-type metric in the
asymptotic region:\footnote{By an AdS$_D$-type metric, we mean one
with the same general structure as AdS$_D$ described in Poincar\'e
coordinates, except that the constant-radius sections carry a curved
metric, rather than a Minkowski metric, as, for example, in (\ref{typem}).}
\be
ds_D^2 = dz^2 + e^{2z}\, (-dt^2 + dx^i\, dx^i + d\rho^2 +
e^{2\gamma'\, \rho}\, dx^2)\,. \label{typem}
\ee
In other words, asymptotically the geometry of AdS$_D$ type,
with a Minkowski$_{D-3}\times H^2$ boundary.  In this case,
the interpolation coordinate is $z$.  Alternatively, we can use
the other radial coordinate for interpolation, namely $\rho$. In
this case, the natural guess for the asymptotic behavior must
again be an AdS$_D$-type metric but with the boundary
AdS$_{D-2}\times S^1$.  The metric is of the form
\be ds_D^2 = d\rho^2 + e^{2\rho}\, \Big(dx^2 + dz^2 + e^{2\wtd\gamma'\,
z}\, (-dt^2 + dx^i\, dx^i)\Big)\,. \ee
We find that such a solution indeed exists and is closely related
to the AdS black hole.  This can be seen by looking at the metric ansatz
for these solutions:
\be ds_D^2 = a^2\, dx^2 + d\rho^2 + c^2\, ds^2_{\rm{AdS}_{D-2}}\,,
\label{metans}
\ee
which is simply the Wick rotation of the AdS black hole ansatz:
\be ds_D^2 =-\td a^2\, dt^2 + d\rho^2 + \td c^2\,
d\Omega_{D-2}^2\,. \label{bhans}\ee
The resulting magnetic solution is called a flux-brane because the
supporting $U(1)$ field strength is proportional to the volume of
the whole transverse space spanned by the radial coordinate $r$
and the angular coordinate $x$.

       The fact that the magnetic flux-brane can be obtained from an
analytical continuation of the standard AdS black hole solution is
not necessarily surprising.  In fact, it was through an analytic continuation 
of this kind that flux-branes were first constructed, in \cite{gibwilt}.
It was observed some time
ago in \cite{gibbons} that a particular limit of a
Reissner-Nordstr\"{o}m black hole can be analytically continued to
the Melvin Universe \cite{melvin}. The latter solution is known as
a flux tube \cite{thorne}, a four-dimensional antecedent of the more
recently-known flux-branes. The recently introduced S-branes
\cite{strom1} and flux-branes \cite{strom2} were constructed
some time ago, in \cite{mukherji1,low} and \cite{mukherji2},
respectively.  In fact, they are obtained by making the same kind of ansatz
as that for standard $p$-branes. Solutions with a rather general
metric ansatz of the form
\be
ds^2=-e^{2U}\, dt^2 + e^{2A}\, d\bar s_q^2 +
e^{2B}\, ds_{\td q}^2\,,\label{oldans}
\ee
were analyzed in detail in \cite{mukherji1}.  In the above metric,
$U,A$ and $B$ are functions of $t$, and $d\bar s_q^2$ and $d\bar
s_{\td q}^2$ denote metrics on maximally-symmetric spaces of
positive, negative or zero curvature, with dimensions $q$ and
$\td q$, respectively. It was shown in \cite{mukherji2} that, with
the appropriate analytical continuation, the ansatz (\ref{oldans})
can give rise to $p$-branes, as well as additional solutions which
have since been called S-branes and flux-branes.

       These S-brane and flux-brane solutions are non-supersymmetric.
Analytical continuations of supersymmetric $p$-branes were obtained in
\cite{clplocal}.  Again, this is not mysterious, since first-order
equations associated with supersymmetry can be viewed as the
``square-root'' of the second-order equations of motion.  Thus, there
can be two different types of first-order equations, providing two
different solutions which are connected by analytical
continuation. Examples of this were given in \cite{clplocal}.

     In the present work, it is pleasantly surprising to find that
the resulting flux-branes in gauged supergravity are both
supersymmetric and regular. Although flux-branes in standard ungauged
supergravities are typically regular, since the flux contribution
vanishes at small distance and hence cannot cause a singularity, they
are generally non-supersymmetric. Not only are our gauged supergravity
solutions regular and supersymmetric, they have the additional
property that the flux-brane world-volume is AdS instead of
Minkowski spaectime.

      The paper is organized as follows.  In section 2, we consider
as a toy model $D$-dimensional anti-de Sitter Einstein-Maxwell gravity.
We give a detailed presentation of the construction of a
non-extremal AdS charged black hole solution using a
superpotential approach. Surprisingly, the extremality condition
does not coincide with the BPS bound required by supersymmetry.
While the supersymmetric AdS black hole is singular, a
supersymmetric and regular magnetic flux-brane solution can be
obtained from analytical continuation.  In section 3, we apply the
same analytical continuation in order to obtain supersymmetric and
regular magnetic flux-branes in AdS gauged supergravities of
dimensions $D=7$, $6$, $5$ and $4$.  In section 4, we find the
$D=5$ first-order equations that follow by requiring supersymmetry.  We show
that the flux-branes solve these equations, and hence that they
are supersymmetric. At first sight, the charge parameters obey a
complicated constraint in order to have small-distance regularity.
However, for particular cases, a more appropriate definition of
charge renders the constraint identical to a simple charge
constraint obeyed by previously-known magnetic brane solutions in
AdS gauged supergravity.  The equivalence of the charge constraint
may hold more generally.  In section 5, we use analytical
continuation to obtain smooth cosmological solutions of de Sitter
Einstein-Maxwell gravity.  The solution interpolates from
dS$_2\times S^{D-2}$ in the infinite past to a dS$_D$-type
spacetime, with an$S^{D-2}\times S^1$ boundary, in the
infinite future.  The conclusions are given in section 6.

\section{AdS Einstein-Maxwell gravity: A toy model}

\subsection{Superpotential and general solution}

         Let us consider $D$-dimensional AdS Einstein-Maxwell theory
with the Lagrangian
\be \hat e^{-1}\, \hat{\cal L}= \hat R - \ft{1}{4}\, \hat F_\2^2
+ (D-1)(D-2)\, g^2\,.\label{emlag}
\ee
The equations of motion for the AdS black hole ansatz of the form
(\ref{bhans}) with $*F_\2 = \lambda\, \Omega_{\sst{(D-2)}}$ can be
summarized in terms of a Hamiltonian $H=T + V$, where $T=\ft12\sum
g_{\alpha\beta}\, \del\varphi^\alpha\, \del\varphi^\beta$,\
$\vec\varphi=(\log c\,, \log a)$ and
\be
g_{\alpha\beta} = \pmatrix{2(D-2)(D-3) & 2(D-2) \cr
                           2(D-2)      & 0}\,.
\ee
$U$ is given by
\be U=a^2\, (-(D-1)(D-2)\, g^2\, c^{2(D-2)}- \epsilon\,
(D-2)(D-3)\, e^{2(D-3)} + \ft12 \lambda^2)\,, \ee
where $\ep=\pm1$ or 0 according to whether the $d\Omega_{D-2}^2$
has positive, negative r zero curvature.  We find that
$U$ can be expressed in terms of a superpotential $W$ as
\be U=-\ft12 g^{\alpha \beta}\,\fft{\partial W}{\partial
\varphi^{\alpha}}\fft{\partial W}{\partial \varphi^{\beta}}\,, \ee
where
\be
W=a\, \Big[4(D-2)^2\,g^2\, c^{2(D-2)}+4\epsilon\,(D-2)^2\,
c^{2(D-3)} + \fft{2(D-2)\,\lambda^2}{D-3} - \alpha\, c^{D-3}
\Big]^{\ft12}\,,
\label{W}
\ee
and $\alpha$ is an arbitrary constant.  From this superpotential,
we can read off the first-order equations
\bea
\dot c &=& \fft{W}{2(D-2)\, a\, c^{D-3}}\,,\nn\\
\dot a &=& \fft{a^2\,\Big(8(D-2)^2\, g^2\, c^{2D} - 4 (D-2)\,
\lambda^2\, c^4 + \alpha\, (D-3)\, c^{D+1}\Big)}{4(D-2)\,
c^{D+2}\, W}\,. \label{emadsbhfo} \eea
The above equations can be solved straightforwardly, by making a
coordinate transformation $d\rho = 2(D-2)a\, c^{D-3}\, W^{-1}\,
dr$, giving
\bea
ds_D^2 &=& - H\, dt^2 + H^{-1}\, dr^2 + r^2\, d\Omega_{D-2}^2\,,\nn\\
{*F_\2}&=&\lambda\, \Omega_{\sst{(D-2)}}\,,\nn\\
H&=&g^2\, r^2\, + \epsilon - \fft{M}{r^{D-3}} +
\fft{Q^2}{r^{2(D-3)}}\,, \label{adsbh} \eea
where the mass $M$ and the charge parameter $Q$ are given by
\be
M=\fft{\alpha}{4(D-2)^2}\,,\qquad
Q=\fft{\lambda}{\sqrt{2(D-2)(D-3)}}\,.
\ee
This is the non-extremal AdS charged black hole solution
\cite{bh1,bh2,bh3,bcs,bh5,tenauthors}. It is rather surprising
that this general non-extremal solution arises from a first-order
system {\it via} a superpotential construction.

\subsection{Supersymmetry Versus Extremality}

        AdS Einstein-Maxwell gravity can be embedded in a
supersymmetric theory for $D=4$ and $D=5$.  We shall show later
that the supersymmetry of the AdS black holes requires the BPS
bound $M=2Q$, as well as $\epsilon=1$.  In this case, the
corresponding superpotential is given by
\be W=2(D-2)\, a\, \sqrt{g^2\, c^{2(D-2)} + (c^{D-3} + Q)^2}\,.
\ee
The solution is then
\be
H=g^2\, r^2 + \Big(1 -\fft{Q}{r^{D-3}}\Big)^2\,.
\ee
Clearly, the solution has a naked singularity at small distance.
For vanishing $g$, $M=2Q$ is precisely the extremality bound for
the black hole.  However, in general, the extremality bound
depends on $g$. We may define a black hole that is extremal by
requiring that there exists a horizon, and furthermore, that
the near-horizon geometry is not of the form $R^2$. Thus, let us
assume that there is a surface located at $r_0$, which satisfies
\be
H(r_0)=0\,,\qquad H'(r_0)=0\,.
\ee
The first equation implies that $r=r_0$ is a horizon, while the
second implies that the geometry is not of the form $R^2$, and
hence has zero temperature.  In fact, the near-horizon geometry is
precisely AdS$_2\times \Omega^{D-2}$.  The solution is given by
\bea ds_D^2&=&-e^{2\gamma\, \rho}\, dt^2 + d\rho^2 + c^2\,
d\Omega_{D-2}^2\,,
\nn\\
{*F_\2} &=&\lambda\, \Omega_{\sst{(D-2)}}\,,
\eea
where $c$ is a non-vanishing constant.  The equations of motion
imply that
\bea
&&(D-1)\, g^2 + \fft{(D-3)\,\epsilon}{c^2} =
\fft{(D-3)\, Q^2}{c^{2(D-2)}}\,,\nn\\
&&\gamma^2=(D-1)\,g^2 + \fft{(D-3)^2\,Q^2}{c^{2(D-2)}}\,.
\label{emads2} \eea
Clearly, there exist solutions for $\epsilon=1$, $-1$ and 0.  The mass
of the solution is given by
\be
M=(\epsilon + g^2\, c^2 + \fft{Q^2}{c^{2(D-3)}})\, c^{D-3}\,.
\ee
Note that in the case of $g=0$ and $\epsilon=1$, the above
condition reduces to $M=2Q$.

\subsection{Supersymmetric and regular flux-branes}

          The fact that the extremality condition does not coincide
with the BPS condition for the AdS black hole is rather
disturbing. One may ask whether the supersymmetric solution is
necessarily singular, or whether on the other hand a smooth extension
can be found.  The answer
turns out to be the latter, and the smooth extension can be
obtained by analytic continuation. All we need do is to
perform Wick rotations such that the $t$ coordinate becomes
spatial and $d\Omega_{D-2}^2$ becomes AdS, dS or Minkowski spacetime.
Since the supersymmetric black hole exists only for $\epsilon=1$,
then in the Wick-rotated solution we must have $\epsilon=-1$, corresponding to
AdS$_{D-2}$. The solution is given by\footnote{For general
solutions, $H=\epsilon + g^2\, r^2 -\fft{M}{r^{D-3}} -
\fft{Q^2}{r^{2(D-3)}}$.}
\bea
ds_D^2 &=&H\, dx^2 + H^{-1}\, dr^2 + r^2\, ds_{\rm{AdS}_{D-2}}^2\,,\nn\\
H&=& g^2\, r^2 - \Big(1 - \fft{Q}{r^{D-3}}\Big)^2\,.
\eea
Clearly, if $Q$ and $g$ are related by
\be
g^2\, Q^{\ft{2}{D-3}} = (D-3)^2\, (D-2)^{-\ft{2(D-2)}{D-3}}\,,
\ee
then we have
\be
H(r_0)=0\,,\qquad H'(r_0)=0\,,
\ee
with $r_0^{D-3}=(D-2)\, Q$.  Thus, the solution interpolates from
AdS$_{D-2}\times H^2$ at small distance ($\rho\rightarrow
-\infty$), given by
\be ds_D^2 = d\rho^2 + e^{2\sqrt{D-2}\, g\,\rho}\, dx^2 + r_0^2\,
ds_{\rm{AdS}_{D-2}}^2\,, \ee
to an AdS$_D$-type geometry, with the boundary AdS$_{D-2}\times
S^1$, in the asymptotic region ($\rho\rightarrow \infty$), given
by
\be ds_D^2 = d\rho^2+e^{2g\, \rho}\, \Big( dx^2 +
ds_{\rm{AdS}_{D-2}}^2\Big)\,. \ee

     The above solution is called a flux-brane, since the 2-form field
strength $F_\2$ is proportional to the volume of the whole transverse
space spanned by the radial coordinate $r$ and the angular coordinate
$x$. Since there is no time component in the field strength, the
solution is magnetic.

     Magnetic branes which interpolate from AdS$_{D-2}\times H^2$ in
the horizon to an asymptotic AdS$_D$-type geometry with the boundary
Minkowski$_{D-3}\times H^2$ have also been found
\cite{sabra1,sorin1,sorin2}. Therefore, AdS$_{D-2}\times H^2$ can flow
to an AdS$_D$-type geometry with two possible boundaries; either
Minkowski$_{D-3}\times H^2$ or AdS$_{D-2}\times S^1$.

\section{Regular supersymmetric flux $(D-3)$-branes}

     In the previous section, we have shown that the AdS
Reissner-Nordstr\"{o}m black hole in $D$-dimensional AdS
Einstein-Maxwell gravity admits an analytical continuation to become a
magnetic flux $(D-3)$-brane.  In four and five dimensions, both
solutions have a BPS limit in which they are supersymmetric, and can
be embedded in a supersymmetric theory.  We now apply the same
analytical continuation in order to obtain multiple-charge magnetic
flux-branes in gauged supergravities of diverse dimensions.

\subsection{$D=7$}

   The bosonic sector of seven-dimensional gauged supergravity,
with a truncation of the gauge fields to the $U(1)\times U(1)$
subgroup that is sufficient for constructing 2-charge black holes,
is described by the Lagrangian
\be
{\cal L}_7 = R\, {*\oneone} - \ft12{*d\vec\phi}\wedge d\vec\phi
 - \ft12 \sum_{i=1}^2 X_i^{-2}\, {*F_\2^i}\wedge
F_\2^i - V\, {*\oneone}\,,\label{d7lagx}
\ee
where
\bea
V &=& \ft12 g^2\, (X_1^{-4}\, X_2^{-4} - 8 X_1\, X_2 - 4 X_1^{-1}\,
X_2^{-2} - 4 X_1^{-2}\, X_2^{-1})\,,\nn\\
X_i &=& e^{-\ft12 \vec a_i\cdot \vec\phi}\,,\qquad
\vec a_1 = (\sqrt2, \sqrt{\ft25})\,,\quad
\vec a_2 = (-\sqrt2, \sqrt{\ft25})\,.
\eea
The two-charge AdS$_7$ black hole in the maximal gauged $D=7$
supergravity was obtained in \cite{tenauthors}. The solution is given by
\bea
ds_7^2 &=& -(H_1\,H_2)^{-4/5}\, f\, dt^2 +
(H_1\,H_2)^{1/5}\, (f^{-1}\, d r^2 + r^2\, d\Omega_{5,k}^2)\ ,\nn\\
f&=& k -\fft{\mu}{r^4} + \ft14 g^2\, r^2\, H_1\, H_2\ ,\qquad
X_i= (H_1\,H_2)^{2/5}\, H_i^{-1}\ ,\nn\\
A_\1^i &=&\sqrt k \, \coth\beta_i\, (1-H_i^{-1})\, dt\ ,\qquad
H_i = 1+ \fft{\mu\, \sinh^2\beta_i}{r^4}\ ,\label{d7adsbh}
\eea
where $d\Omega_{5,k}^2$ is the metric on a unit $S^5$, $T^5$ or $H^5$,
according to whether $k=1,0$ or $-1$.  In particular, we are
interested in the case $k=1$, since only then does there exist an
extremal limit with $ \mu\rightarrow 0$ while keeping $\ell_i^4=
\mu\,\sinh^2\beta_i$ fixed, such that the $A_\1^i$ do not vanish. The
metric and the $X_i$ in the extremal solution have the same form given
in (\ref{d7adsbh}), but with $f=1+\ft14g^2\, r^2\, H_1\, H_2$ and
$H_i=1 + \fft{\ell_i^4}{r^4}$.  The gauge fields are given by
$A_\1^i=(1 - H_i^{-1})\, dt$.  We now perform the following analytical
continuation:\footnote{It is also possible to perform an analytical
continuation such that the sphere transforms into de Sitter
spacetime. However, in that case there is no extremal limit, since
$k=-1$ and hence $\coth\beta_i\rightarrow \cos\beta_i$.}
\be r\rightarrow \im\, r\,,\qquad t\rightarrow x\,,\qquad
f\rightarrow -f \,,\qquad d\Omega_5^2 \rightarrow -
ds_{\rm{AdS}_5}^2\,. \ee
The extremal AdS$_7$ black hole solution becomes an extremal
magnetic flux-brane, given by
\bea
ds_7^2 &=& (H_1\, H_2)^{-4/5}\, f\, dx^2 + (H_1\, H_2)^{1/5}\,
(f^{-1}\, dr^2 + r^2\, ds_{\rm{AdS}_5}^2)\,,\nn\\
f&=& \ft14 g^2\, r^2\, H_1\, H_2 - 1\,,\qquad X_i=H_i^{-1}\,
(H_1\, H_2)^{2/5}\,,\nn\\
A_\1^i&=&(1-H_i^{-1})\, dx\,,\qquad
H_i=1 + \fft{\ell_i^4}{r^4}\,.\label{d7fluxb}
\eea
Note that $t$ automatically becomes spacelike without a Wick
rotation. On the other hand, although we have sent $r$ to $\im\, r$, the
coordinate $r$ remains spatial.  The metric approaches an AdS$_7$-type
metric with an AdS$_5\times S^1$ boundary in the asymptotic region
$r\rightarrow \infty$, given by
\be
ds_7^2 = r^2\,(\ft14 g^2\, dx^2 + ds_{\rm{AdS}_5}^2)  +
\fft{4dr^2}{g^2\,r^2}\,.
\ee
In general, the metric is singular at small distance. However, for
appropriate choices of the charge parameters $\ell_i$, it becomes
AdS$_5\times H^2$ at small distance, and hence gives rise to a
smooth solution interpolating from AdS$_5\times H^2$ at small
distance to the AdS$_7$-type metric in the asymptotic region.
Clearly, the condition for such a solution is that there exist an
$r_0$ such that
\be
f(r_0)=0\,,\qquad f'(r_0)=0\,.\label{fcons}
\ee
Eliminating $r_0$ in the above two equations provides a relation
between $\ell_i$ and $g$ for which the metric becomes AdS$_5\times
H^2$ as $r$ approaches $r_0$. For the general solution
(\ref{d7fluxb}), the constraints (\ref{fcons}) become
\be
g^2\, (r_0^4 +\ell_1^4)\, (r_0^4 + \ell_2^4) -4r_0^6=0\,,\qquad
r_0^8-(\ell_1^4 + \ell_2^4)\,r_0^4 - 3\ell_1^4\, \ell_2^4=0\,.
\ee

    We shall now consider special cases. For $\ell_1=\ell$ and
$\ell_2=0$, we have
\be r_0=\ell\,,\qquad g^2\,\ell^2=2\,, \ee
and the corresponding function $f$ is given by
\be f=\fft{(r^2-\ell^2)^2}{2\ell^2\, r^2}\,. \ee
Thus, the metric clearly approaches AdS$_5\times H^2$ near $r_0$.
Another simple case is when $\ell_1=\ell_2=\ell$, for which we
have
\bea
r_0&=&3^{1/4}\,\ell\,,\qquad
g^2\,\ell^2 = \ft34\,\sqrt3\,,\nn\\
f&=&\fft{(r^2-r_0^2)^2\, \Big[(3r^2 + r_0^2)^2 - 4r_0^2\, r^2\Big]}{
16r_0^2\,r^6}\,.
\eea
The geometry again approaches AdS$_5\times H^2$ as $r\rightarrow
r_0$.

        For generic $\ell_i$, we find after some algebra that the
requirement of an asymptotic AdS$_5\times H^2$ form is achieved when the
parameters satisfy the real solution of the equation
\be g^8\,(\ell_1^4-\ell_2^4)^4 - 4g^4\,(\ell_1^4 + \ell_2^4)(
\ell_1^8 - 34\ell_1^4\,\ell_2^4 + \ell_2^8) - 432
\ell_1^4\,\ell_2^4 =0\,. \ee

It is straightforward to lift the general solution (\ref{d7fluxb}) to
$D=11$ by using the reduction ansatz obtained in
\cite{tenauthors}. The M-theory metric is given by
\bea ds_{11}^2 &=& \Delta^{\ft13}\, \Big[r^2\,ds_{\rm{AdS_5}}^2+
f^{-1}dr^2+(H_1\,H_2)^{-1}f\,dx^2\nn\\
& & +\fft{1}{g^2\,\Delta} \Big\{ d\mu_0^2+\sum_{i=1}^2H_i\,\Big(
d\mu_i^2+\mu_i^2\, (d\phi_i +g\,
(1-H_i^{-1})\,dx)^2\Big)\Big\}\Big]\,, \label{7to11} \eea
where $\mu_i$ are spherical coordinates which satisfy $\mu_0^2 +
\mu_1^2 + \mu_2^2=1$.  The warp factor $\Delta$ is given by
\footnote{$\Delta=(H_1\,H_2)^{\ft35}\,\td \Delta$, where $\td
\Delta$ is the warp factor defined in \cite{tenauthors}.}
\be \Delta=H_1H_2\,\mu_0^2 + H_2\,\mu_1^2 + H_1\,\mu_2^2\,. \ee
Since $\Delta>0$ and $r\ge r_0>0$, the metric (\ref{7to11}) has no
power-law singularity.

The metric (\ref{7to11}) is a warped product of AdS$_5$ with an
internal metric which can be regarded as having a reduced generalized
holonomy group, since it is not Ricci-flat and involves a form
field. As $r$ approaches $r_0$, the internal space can be viewed as an
$S^4$ bundle over $H^2$, with two diagonal $U(1)$
fibres\footnote{Supersymmetric AdS solutions of M-theory with internal
spaces that can be viewed as $S^n$ bundles over various spaces have
also been discussed in
\cite{cham,nunez,gauntlett1,gauntlett2,nunez2,gauntlett3,sorin1,sorin2}.}.

\subsection{$D=6$}

   With appropriate matter fields, the bosonic sector of
six-dimensional gauged supergravity can be truncated to a $U(1)\times
U(1)$ gauge subgroup, which is sufficient for constructing 2-charge
black holes.  The corresponding Lagrangian is given by
\be {\cal L}_6 = R\, {*\oneone} - \ft12{*d\vec\phi}\wedge
d\vec\phi
 - \ft12 \sum_{i=1}^2 X_i^{-2}\, {*F_\2^i}\wedge
F_\2^i - V\, {*\oneone}\,,\label{d6lagx} \ee
where
\bea V &=& \ft49 g^2\, (X_0^2 - 9 X_1\, X_2 - 6 X_0\,
X_1 - 6 X_0\, X_2)\,,\nn\\
X_i &=& e^{-\ft12 \vec a_i\cdot \vec\phi}\,,\qquad \vec a_1 =
(\sqrt2, \fft{1}{\sqrt2})\,,\quad \vec a_2 = (-\sqrt2,
\fft{1}{\sqrt2})\,, \eea
and $X_0\equiv (X_1\, X_2)^{-3/2}$.

     The AdS$_6$ black hole was obtained in \cite{d6gauged}.  Using
the same convention and notation of \cite{d6gauged} and \cite{gub},
we can follow the same strategy and obtain the extremal $D=6$
flux-branes, given by
\bea
ds_6^2 &=& (H_1\, H_2)^{-3/4}\, f\, dx^2 + (H_1\, H_2)^{1/4}\, (
f^{-1}\, dr^2 + r^2\, ds_{\rm{AdS}_4}^2)\,,\nn\\
A_\1^i&=&(1-H_i^{-1})\,dx\,,\qquad X_i=(H_1\,H_2)^{3/8}\, H_i^{-1}\,,\nn\\
f&=&\ft29\,g^2\,r^2 H_1\,H_2 -1\,,\qquad H_i=1 +
\fft{\ell_i^3}{r^3}\,. \label{d6fluxb} \eea
Analogous to the $D=7$ example, the solution becomes AdS$_6$-type
with the boundary of AdS$_4\times S^1$.  At small distance, the
solution can be regular if there exists an $r_0$ which satisfies
the constraint (\ref{fcons}).

Let us consider two simple cases.  For $\ell_1=\ell$ and
$\ell_2=0$, we find that
\bea
r_0^3=\ft12\ell^3\,,\qquad 2g^2\,\ell^2=3\,\, 2^{2/3}\,,\qquad
f=\fft{(r-r_0)^2\, (r+2r_0)}{2r_0^2\, r}\,.
\eea
A second case is when $\ell_1=\ell_2=\ell$, for which we find that
\bea
r_0^3&=& 2\ell^3\,,\qquad
g^2\,\ell^2=2^{1/3}\,,\nn\\
f&=& \fft{(r-r_0)^2(2r + r_0)(2r^3 + 3r_0\, r^2 + r_0^3)}{9
r_0^2\, r^4}\,,
\eea
For both cases, the metric approaches AdS$_4\times H^2$ as $r$
approaches $r_0$.

    For generic $\ell_i$, the constraint is given by
\be g^{12}\, (\ell_1^3-\ell_2^3)^6 - 27 g^6\, \Big( \ell_1^{12} +
\ell_2^{12} - 282\ell_1^6\,\ell_2^6 - 76\ell_1^3\,\ell_2^3\,
(\ell_1^6 + \ell_2^6)\Big) - 23328\ell_1^3\,\ell_2^3=0\,. \ee

It is straightforward to lift the general solution (\ref{d6fluxb}) to
$D=10$ by using the reduction ansatz obtained in
\cite{d6gauged,gub}. The ten-dimensional metric is given by
\bea ds_{10}^2 &=& \mu_0^{\ft{1}{12}}\,\Delta^{\ft38}\,
\Big[r^2\,ds_{\rm{AdS_4}}^2+
f^{-1}dr^2+(H_1\,H_2)^{-1}f\,dx^2\nn\\
& & +\fft{1}{g^2\,\Delta} \Big\{ d\mu_0^2+H_1\,\Big(
d\mu_1^2+\mu_1^2\, (d\phi_1 +g\, (1-H_1^{-1})\, dx)^2\Big)\nn\\
&& +H_2\,\Big( d\mu_2^2+\mu_2^2\,
(d\phi_2+g\, (1-H_2^{-1})\,dx)^2\Big) \Big\}\Big],
\label{6to10} \eea
where $\mu_i$ are spherical coordinates which satisfy $\mu_0^2 +
\mu_1^2 + \mu_2^2=1$.  The warp factor $\Delta$ is given by
\be \Delta=H_1H_2\,\mu_0^2 + H_2\,\mu_1^2 + H_1\,\mu_2^2\,. \ee
Note that the metric (\ref{6to10}) is singular, due to the overall
angular factor.

\subsection{$D=5$}

  In the conventions and notation of \cite{tenauthors}, the relevant
bosonic sector of five-dimensional gauged supergravity, truncated to
the $U(1)^3$ subgroup of $SO(6)$, is described by the Lagrangian
\be
{\cal L}_5 = R\, {*\oneone} -\ft12 {*d\vec\phi}\wedge d\vec\phi
- \ft12 \sum_{i=1}^3 X_i^{-2}\, {*F_\2^i}\wedge F_\2^i + \ft1{6}
\ep_{ijk}\, F_\2^i\wedge F_\2^j\wedge A_\1^k
- V\, {*\oneone}\,,
\ee
where
\bea
V &=& - 4 g^2\, \sum_{i=1}^3 X_i^{-1}\,,\\
X_i &=& e^{-\ft12\vec a_i\cdot\vec\phi}\,,\qquad \vec a_1 =
(\ft{2}{\sqrt6}, \sqrt2)\,,\quad \vec a_2 = (\ft{2}{\sqrt6},
-\sqrt2)\,,\quad \vec a_3 = (-\ft{4}{\sqrt6},0)\,.\nn \eea
The three-charge AdS$_5$ black hole was obtained in \cite{bh1,bcs}.
Following the same procedure as we used previously, we obtain AdS$_5$
flux-branes given by
\bea
ds_5^2 &=& (H_1\, H_2\, H_3)^{-2/3}\, f\, dx^2 +
(H_1\, H_2\, H_3)^{1/3}\, (f^{-1}\, dr^2 + r^2\, ds_{\rm{AdS}_3}^2)
\,,\nn\\
X_i&=& H_i^{-1}\, (H_1\, H_2\, H_3)^{1/3}\,,\qquad
A_\1^i=(1- H_i^{-1})\, dx\,,\nn\\
f&=& g^2\, r^2\, (H_1\, H_2\, H_3) - 1\,, \qquad H_i = 1 +
\fft{\ell_i^2}{r^2}\,. \label{d5fluxb} \eea
Asymptotically, the metric becomes AdS$_5$-type, with an
AdS$_3\times S^1$ boundary.  In general, the solution is singular at small
distance.  However, we are interested in the possibility of
finding choices of $\ell_i$ such that the small-distance behavior
approaches AdS$_3\times H^2$.  This can be achieved by satisfying the
condition (\ref{fcons}).  The elimination of $r_0$ in the two
equations gives rise to a constraint on $\ell_i$ and $g$.

        If there is only one non-vanishing $\ell_i$, then it can be
easily seen that the constraint (\ref{fcons}) cannot be satisfied.
Hence in this case, there can be no AdS$_3\times H^2$ limit.  This
was also observed in \cite{sorin1,sorin2}.  Now let us consider
$\ell_3=0$, with non-vanishing $\ell_1$ and $\ell_2$. We find that
\bea r_0^2=\ell_1\,\ell_2\,,\qquad g\,(\ell_1 + \ell_2)=1\,,\qquad
f=\fft{(r^2-r_0^2)^2}{r^2(\ell_1+\ell_2)^2} \,. \eea
Another simple case arises when $\ell_i=\ell$ for all $i$, in
which case
\bea r_0^2=2\ell^2\,,\qquad g^2\,\ell^2 = \ft4{27}\,,\qquad f=
\fft{(r^2-r_0^2)^2\, (8r^2 + r_0^2)}{27 r_0^2\,r^4}\,. \eea
For both cases, the metric approaches AdS$_3\times H^2$ as $r$
approaches $r_0$.

         For generic $\ell_i$, the constraint is given by
\bea
&& g^6 (\ell_1^2-\ell_2^2)^2 (\ell_2^2-\ell_3^2)^2
(\ell_3^2-\ell_1^2)^2 + g^4\,\Big(-2\ell_1^2\,\ell_2^2\,
(\ell_1^4 + \ell_2^4) + \hbox{cyclic on 1,2,3}\nn\\
&&+4\ell_1^2\,\ell_2^2\,\ell_3^2\,(2\ell_1^4 + 2\ell_2^4 +
2\ell_3^4 - \ell_1^2\,\ell_2^2 - \ell_2^2\,\ell_3^2 -
\ell_3^2\,\ell_12)\Big)\nn\\
&&+g^2\,\Big( \ell_1^4\,\ell_2^4 + \ell_2^4\,\ell_3^4 +
\ell_3^4\,\ell_1^4 -10 \ell_1^2\,\ell_2^2\,\ell_3^2\,( \ell_1^2 +
\ell_2^2 + \ell_3^2)\Big) + 4\ell_1^2\,\ell_2^2\,\ell_3^2 =0\,.
\label{5d3ccon} \eea

It is straightforward to lift the general solution (\ref{d5fluxb})
to $D=10$ by using the reduction ansatz obtained in \cite{tenauthors}. The
ten-dimensional metric is given by
\bea ds_{10}^2 &=& \sqrt{\Delta}\, \Big[r^2\,ds_{\rm{AdS_3}}^2+
f^{-1}dr^2+(H_1\,H_2\,H_3)^{-1}f\,dx^2\nn\\
& & +\fft{1}{g^2\,\Delta}\,\sum_{i=1}^3H_i\,\Big(
d\mu_i^2+\mu_i^2\, (d\phi_i+g\,
(1-H_i^{-1})\,dx)^2\Big)\Big]\,, \label{5to10} \eea
where $\mu_i$ are spherical coordinates which satisfy $\sum_i
\,\mu_i^2=1$. The warp factor $\Delta$ is given by
\be \Delta=H_1\,H_2\,H_3 \sum_{i=1}^3 H_i^{-1}\,\mu_i^2\,. \ee
Since $\Delta>0$ and $r\ge r_0>0$, the metric (\ref{5to10}) has no
power-law singularity.

\subsection{$D=4$}

   In the conventions and notation of \cite{tenauthors}, the relevant
bosonic sector of the $U(1)^4$ truncation of gauged $SO(8)$
four-dimensional supergravity is described by the Lagrangian
\be
{\cal L}_4 = R\, {*\oneone} - \ft12 {*d\vec\phi}\wedge d\vec\phi -
\ft12 \sum_{i=1}^4\, X_i^{-2}\, {*F_\2^i}\wedge F_\2^i - V\, {*\oneone}
\,,
\ee
where
\bea
V &=& -2 g^2 \, \sum_{i,j} X_i\, X_j\,,\qquad X_i = e^{-\ft12\vec a_i
\cdot \vec\phi}\,,\\
\vec a_1 &=&(1,1,1)\,,\quad \vec a_2 = (1,-1,-1)\,,\quad
\vec a_3 =(-1,1,-1)\,,\quad \vec a_4 = (-1,-1,1)\,.\nn
\eea
The four-charge AdS black hole in $D=4$ maximal gauged supergravity
was obtained in \cite{bh5}.  Following the same procedure we discussed
earlier, we can obtain magnetic flux-branes, given by
\bea
ds_4^2 &=& (H_1\, H_2\, H_3\,H_4)^{-1/2}\, f\, dx^2 +
(H_1\, H_2\, H_3\,H_4)^{1/2}\, (f^{-1}\,dr^2 + r^2\,
ds_{\rm{AdS}_2}^2)\,,\nn\\
A_\1^i&=&(1-H^{-1}_i)\,dx\,,\qquad X_i= H_i^{-1}\,
(H_1\, H_2\, H_3\,H_4)^{1/4}\,,\nn\\
f&=&4g^2\,r^2\, (H_1\, H_2\, H_3\,H_4) -1\,,\qquad H_i=1 +
\fft{\ell_i}{r}\,. \label{d4fluxb} \eea
Asymptotically, the metric approaches AdS$_4$-type, with an
AdS$_2\times S^1$ boundary.  With an appropriate choice of the parameters
$\ell_i$, the metric approaches AdS$_2\times H^2$ at small distance.
The constraint for general $\ell_i$ is given by (\ref{fcons}). Let
us look at a few examples.  First, if more than one
$\ell_i$ vanishes, the solution does not admit an
AdS$_2\times H^2$ limit. For the simple choice $\ell_3=\ell_1$
and $\ell_4=\ell_2$, we have
\bea
r_0^2&=&\ell_1\,\ell_2\,,\qquad
2g\, (\sqrt{\ell_1} + \sqrt{\ell_2})^2 =1\,,\nn\\
f&=&\fft{(r-r_0)^2\,\Big( (r-r_0)^2 + 2r\, (\sqrt{\ell_1} +
\sqrt{\ell_2})^2 \Big)}{(\sqrt{\ell_1} + \sqrt{\ell_2})^4\, r^2}
\,.
\eea
Another simple example is provided by $\ell_1=\ell_2=\ell_3=\ell$
and $\ell_4=0$, in which case we have
\be r_0=\ft12\ell\,,\qquad 27g^2\,\ell^2=1\,,\qquad
f=\fft{(2r-\ell)^2\, (r+4\ell)}{27\ell^2\, r}\,. \ee
For the case of all $\ell_i$ equal, the system reduces to the
Einstein-Maxwell theory discussed in section 2. For these cases,
the metric approaches AdS$_2\times H^2$ as $r$ approaches $r_0$.

      It is straightforward to lift the general solution
(\ref{d4fluxb}) to $D=11$ by using the reduction ansatz obtained in
\cite{tenauthors}. The M-theory metric is given by
\bea ds_{11}^2 &=& \Delta^{\ft23}\, \Big[r^2\,ds_{\rm{AdS_2}}^2+
f^{-1}dr^2+(H_1\,H_2\,H_3\,H_4)^{-1}f\,dx^2\nn\\
& & +\fft{1}{g^2\,\Delta}\,\sum_{i=1}^4H_i\,\Big(
d\mu_i^2+\mu_i^2\, (d\phi_i+g\,
(1-H_i^{-1})\,dx)^2\Big)\Big]\,, \label{4to11} \eea
where $\mu_i$ are spherical coordinates which satisfy
$\sum_i\,\mu_i^2=1$. The warp factor $\Delta$ is given by
\be \Delta=H_1\,H_2\,H_3\,H_4 \sum_{i=1}^4 H_i^{-1}\,\mu_i^2\,.
\ee
Since $\Delta>0$ and $r\ge r_0>0$, the metric (\ref{4to11}) has no
power-law singularity.

\section{Supersymmetry and first-order equations}

\subsection{Supersymmetry analysis for $D=5$}

   We shall consider the case of $D=5$ as an example.
   The supersymmetry transformations of the fermionic fields in the
truncated $U(1)^3$ gauged five-dimensional supergravity are
given by
\bea
\delta\lambda_\a &=& \Big(\ft18 \fft{\del X_i^{-1}}{\del \phi_\a}\,
\Gamma^{MN}\, F^i_{MN} - \ft{\im}4\, \Gamma^M\, \del_M\phi_\a
 + \ft{\im}{2}g\, \sum_{i=1}^3 \fft{\del X_i}{\del\phi_\a}\, \Big)\,\ep
\,,\label{d5susytrans}\\
\delta\psi_M &=& \Big(\nabla_M + \ft{\im}{24} X_i^{-1}\, F^i_{NP}\,
(\Gamma_M{}^{NP} - 4 \delta_M^N\, \Gamma^P) + \ft16 g\, \sum_{i=1}^3
X_i\, \Gamma_M - \ft{\im}2\, g\, \sum_{i=1}^3 A^i_M\Big)\, \ep\,.
\nn
\eea

    We take the metric ansatz
(\ref{metans}) and write the three 1-form potentials as
\be
A^i = - u_i\, dx\,,
\ee
where the $u_i$ are functions only of $r$.  From (\ref{d5susytrans}),
we find that the first-order equations
following from requiring supersymmetry are
\bea
\dot\phi_1^2\!\!\! &=&\!\!\!
\ft23 g^2\, (X_1 + X_2 - 2 X_3)^2 - \fft1{6a^2}\,
(\dot u_1\, X_1^{-1} + \dot u_2\, X_2^{-1} - 2 \dot u_3\, X_3^{-1})^2
\,,\nn\\
\dot\phi_2^2\!\!\! &=&\!\!\! 2 g^2\, (X_1 - X_2)^2 - \fft1{2a^2}\,
(\dot u_1\, X_1^{-1} - \dot u_2\, X_2^{-1})^2
\,,\nn\\
\dot a^2\!\!\! &=&\!\!\!
\ft19 a^2\, g^2\, (X_1 +X_2 +X_3)^2 + g^2\, (u_1+u_2+
u_3)^2  - \ft19 (\dot u_1\, X_1^{-1} + \dot u_2\, X_2^{-1}
           + \dot u_3\, X_3^{-1})^2\,,\nn\\
\dot c^2\!\!\! &=&\!\!\!  \ft19 c^2\, g^2\, (X_1 +X_2 +X_3)^2
         -[-1 + \fft{c}{6a}\, (\dot u_1\, X_1^{-1} + \dot u_2\, X_2^{-1}
           + \dot u_3\, X_3^{-1})]^2\,.\label{d5fo}
\eea
In addition, the field equations for the three vector potentials imply
\be \dot u_1 = \fft{a\, X_1^2\, q_1}{c^3}\,,\qquad \dot u_2 =
\fft{a\, X_2^2\, q_2}{c^3}\,,\qquad \dot u_3 = \fft{a\, X_3^2\,
q_3}{c^3}\,, \label{d5u} \ee
where the $q_i$ are constants.  It is straightforward to verify
that the our solution (\ref{d5fluxb}) satisfies these first-order
equations and hence it is supersymmetric.  The constants $q_i$ are
given by $q_i=2\ell_i^2$.  Note that, when all $q_i$ are equal, we
can set $X^i=1$ and the system reduces to Einstein-Maxwell gauged
supergravity.  It is straightforward to verify that supersymmetry
implies that $M=2Q$, as stated in section 2.2.

     It is worth mentioning that the first-order equations
(\ref{d5fo}) and (\ref{d5u}) do not necessarily imply the
second-order equations of motion.  To illustrate this, let us
consider a simpler case with $u_1=u_2=0$ and $u_3$ non-vanishing.
In this case, we can set $\phi_2=0$, which reduces the equations
to a single-charge and single-scalar system. Substituting the
first-order equations into the equations of motion, we find that
an additional algebraic equation has to be satisfied, namely
\be
c^2=\fft{q_3\, e^{\ft2{\sqrt6}\,\phi_1}}{2(1-e^{\ft3{\sqrt6}\,\phi_1})}
\,.\label{d5newcon}
\ee
It is easy to see that this constraint is consistent with the
first-order equations.  In the special case of $u_1=u_2=0$ and
$\phi_2=0$, the $\dot\phi_1$ and $\dot c$ equations of
(\ref{d5fo}) imply the following relation:
\be q_3\,b^8\,C'^2 - (q_3\,b^2-6 C)^2\,b^2\,C - 4g^2\, C^3\,
(b^2\,(b^3-1)^2\, C'^2 - (b^3+2)^2\, C^2)=0\,, \label{d5cb} \ee
where $C=c^2$, $b=e^{\phi_1/\sqrt6}$ and $C'\equiv dC/db$.  The
constraint (\ref{d5newcon}) is a $g$-independent solution to
(\ref{d5cb}).  While (\ref{d5cb}) clearly admits a more general
class of solutions with an integration constant, only the special
solution (\ref{d5newcon}), with a specific integration constant, is
consistent with the second-order equations of motion.  This
provides an example in which the Killing spinor equations do not
automatically give rise to a solution of the equations of motion.  The fact
that the constraint (\ref{d5newcon}) is independent of $g$ implies
that the solution is supersymmetric for both $g=0$ as well as
non-vanishing $g$. (Of course, setting $g=0$ requires appropriate
Wick rotations of coordinates, after which the solution becomes an
electric black hole in ungauged supergravity.)

       Recall that in section 2 we obtained first-order equations
with a superpotential construction for the AdS black holes and
flux-branes of AdS Einstein-Maxwell theory.  We found that the
superpotential approach could not be extended to cases where scalar fields are
present in the gauged supergravity theories.  This can be
understood from the fact that the first-order system is then not of itself
enough to determine the solution; additional algebraic equations are
needed in these cases, which are not expected to arise from
a superpotential description.

          While $q_i=2\ell_i^2$ are the charge parameters of the AdS
black hole solutions, in the related AdS flux-branes it is no longer
appropriate to view these as the natural charge parameters.
To see this, let us consider the case in which
(\ref{metans}) is the metric on AdS$_3\times H^2$, for which $a=e^{\gamma\,
\rho}$ and $c$ is a constant. The metric can then be expressed as
\be ds_5^2=\gamma^{-2}\, d\Omega_2^2 + c^2\, ds_{{\rm AdS}_3}^2\,,
\ee
where $d\Omega_2$ and $ds_{{\rm AdS}_3}^2$ are the metrics of the
unit $H^2$ and AdS$_3$, respectively.  Then we have
\be dA_\1^i=-\dot u_i\, d\rho\wedge dx= -\fft{\dot
u_i}{a\,\gamma^2}\, \Omega_\2 \equiv \td q_i\,\Omega_\2\,, \ee
where $\td q_i$ is the charge as defined in \cite{sorin1,sorin2}.
Thus, we find that
\be
\td q_i=\fft{X_i^2\,q_i}{c^3\,\gamma^2}\,.
\ee
We expect that the complicated expression (\ref{5d3ccon}) becomes
the same as the constraint found in \cite{sorin1,sorin2}, once expressed in
terms of the parameters $\td q_i$.  As an illustration, let us consider the
``two-charge'' case, with $\ell_1$ and $\ell_2$ non-vanishing, but
$\ell_3=0$.  The AdS$_3\times H^2$ solution can easily be found from the
results in section 3.3, leading to
\be
\td q_1=\td q_2=\ft12 (\ell_1 + \ell_2)=\ft12 g^{-1}\,.
\ee
This is precisely the same as the condition given in
\cite{sorin1,sorin2}. It is rather surprising that the
two-charge solutions with two ostensibly different charges in fact have
equal charges in the more natural parameterisation.

\subsection{Supersymmetry in $D=7$}

    As a further example, we consider the supersymmetry of the $D=7$
solutions.  The supersymmetry transformations of the fermionic fields
in the truncated $U(1)\times U(1)$ gauged seven-dimensional
supergravity are given by
\bea
\delta\lambda_1\!\!\! &=&\!\!\!
\Big(-\ft1{16} X_1^{-1}\, F^1_{MN}\, \Gamma^{MN}\,
\sigma_{12} -(\fft{3\del_M X_1}{4X_1} + \fft{2\del_M X_2}{4X_2})\,
\Gamma^M
 + \ft14 g\, (X_1 - X_1^{-2}\, X_2^{-2})\Big)\, \ep\,,\nn\\
\delta\lambda_2\!\!\! &=&\!\!\!
\Big(-\ft1{16} X_2^{-1}\, F^2_{MN}\, \Gamma^{MN}\,
\sigma_{34} -(\fft{2\del_M X_1}{4X_1} + \fft{3\del_M X_2}{4X_2})\,
\Gamma^M
   + \ft14 g\, (X_2 - X_1^{-2}\, X_2^{-2})\Big)\, \ep\,,\nn\\
\delta\psi_M\!\!\! &=&\!\!\! \Big(\nabla_M + \ft14(X_1^{-1}\, F^1_{MN}\,
\sigma_{12} + X_2^{-1}\, F^2_{MN}\, \sigma_{34})\, \Gamma^N
+ \ft14 g\, X_1^{-2}\, X_2^{-2} \,\Gamma_M
\nn\\
\!\!\!&&\!\!\!+ \ft14 \Gamma_M\, \Gamma^N\, (X_1^{-1}\,
\del_N X_1 + X_2^{-1}\,
\del_N X_2) +
\ft12 g\, (A^1_M\, \sigma_{12} + A^2_M\, \sigma_{34})\Big)\, \ep\,,
\eea
where $\sigma_{12}$ and $\sigma_{34}$ are generators in the Cartan
subalgebra of $SO(5)$.

    From these, and writing $A^i = - u_i\, dx$ as before, we find that
the requirements of supersymmetry imply the following first-order
equations:
\bea
\Big(\fft{3 \dot X_1}{X_1} + \fft{2 \dot X_2}{X_2}\Big)^2 &=&
      4 g^2 \, (X_1 - X_1^{-2}\, X_2^{-2})^2 -
             \fft{\dot u_1^2}{a^2\, X_1^2}\,,\nn\\
\Big(\fft{2\dot X_1}{X_1} + \fft{3 \dot X_2}{X_2}\Big)^2 &=&
      4 g^2 \, (X_2 - X_1^{-2}\, X_2^{-2})^2 -
             \fft{\dot u_2^2}{a^2\, X_2^2}\,,\nn\\
\Big(\fft{2\dot a}{a} + \fft{\dot X_1}{X_1} + \fft{\dot
X_2}{X_2}\Big)^2 &=&
g^2\, X_1^{-4}\, X_2^{-4} - \fft1{a^2}\,
\Big(\fft{\dot u_1}{X_1} + \fft{\dot u_2}{X_2}\Big)^2 +
\fft{4 g^2\, (u_1+u_2)^2}{a^2}\,,\nn\\
\Big(\fft{2\dot c}{c} + \fft{\dot X_1}{X_1} + \fft{\dot
X_2}{X_2}\Big)^2 &=& g^2\, X_1^{-4}\, X_2^{-4} - \fft{4}{c^2}\,.
\eea
Additionally, the field equations for the gauge fields imply
\be
\dot u_1 = \fft{a\, X_1^2\, q_1}{c^5}\,,\qquad
\dot u_2 = \fft{a\, X_2^2\, q_2}{c^5}\,,
\ee
where $q_i$ are constants.  Our $D=7$ flux-branes satisfy these
first-order equations and hence are supersymmetric.

\section{Analytical continuation to de Sitter spacetime}

      A large class of supersymmetric magnetic brane solutions
supported by $U(1)$ gauge fields in AdS gauged supergravities have
previously been investigated
\cite{romans,sabra3,sabra4,nunez,gauntlett4,sabra1,sabra2,klemm,%
sorin1,sorin2}.  These solutions smoothly interpolate between
AdS$_{D-2}\times \Omega^2$ (where $\Omega^2=S^2$ or $H^2$) at the
horizon and an AdS$_D$-type geometry in the asymptotic region, for
$4\le D\le 7$.  The boundary geometry of the AdS$_D$-type metric is
Minkowski$_{D-3}\times \Omega^2$.

      De Sitter Einstein-Maxwell theory and, in a supergravity
setting, Hull's exotic * theories which have flux kinetic terms with
the ``wrong'' sign \cite{hull}, support the de Sitter counterpart
to these types of solutions.  These are smooth cosmological
solutions in which the proper time runs from an infinite past
that is dS$_{D-2}\times S^2$ to an infinite future that is a
dS$_D$-type spacetime with an $R^{D-3}\times S^2$ boundary
\cite{dscos}. These cosmological solutions can be obtained
directly from the aforementioned magnetic brane solutions of
standard gauged supergravity via analytical continuation.

      An analogous analytical continuation can apply to our
flux-branes. However, there is one important difference. As we
shall see, the resulting $U(1)$ gauge fields still have
standard kinetic terms, with the usual sign. The
analytical continuation can be implemented by the replacements
\be
g\rightarrow \im\, g\,,\qquad r\rightarrow \im\, t\,,\qquad
\ell_i\rightarrow \im\,\ell_i\,,\qquad
ds_{\rm{AdS}_d}^2 \rightarrow -ds_{S^d}^2\,.
\ee
Under these transformations, the functions $H_i$ become
\be H\rightarrow H=1 + \fft{\ell_i^{d-1}}{t^{d-1}}\,, \ee
and hence the gauge fields $A_\1^i = (1 - H_i^{-1})\, dx$ remain
real. It follows that the signs of the corresponding kinetic terms
remain unaltered.  The function $f$ and the scalars $X_i$
maintain the same form as in the flux-branes. Since the forms
of these cosmological solutions are more or less the same as the
corresponding flux-branes, we shall not repeat them here.

      The transformation $g\rightarrow \im\, g$ implies that the the
cosmological solution is asymptotic to a dS-type geometry at infinity,
with an $S^{d}\times S^1$ boundary, rather than to an AdS-type
geometry. Since the kinetic terms for the $U(1)$ gauge fields have the
standard sign, it follows that our flux-brane solutions exist in
standard de Sitter supergravity theories instead of those associated
with the * theories. Cosmological solutions of such theories in $D=5$
and $D=4$ supported by only the scalar potential were studied in
\cite{bc}.  The solution are typically singular at certain point in
the past.  Our cosmological solutions however can be totally regular.
While it is not clear how our scalar-coupled systems with
$g\rightarrow \im\, g$ could be obtained from string or M-theories, de
Sitter Einstein-Maxwell gravity in $D=5$ or $D=4$ can arise from type
IIB and M-theory, respectively.  Here, we shall present non-trivial
cosmological solutions to such theories obtained from first-order
equations, even though supersymmetry is not expected.

       Note that the cosmological solution can be obtained directly
from the AdS black hole solutions, by sending $g$ to $\im\,g$, in
which case, the coordinate $r$ becomes automatically timelike without
the Wick rolation.  There is an ``interior'' region where the
coordinate $r$ is spatial.  In this region the solution is singular.
This is analogous to the Schwarzchild black hole, which behaves like a
cosmological solutino inside the horizon.  The difference is that in
our solution, the region with $r$ being time-like is totally regular.

    The Lagrangian for de Sitter Einstein-Maxwell gravity is given by
\be
e^{-1}\, {\cal L}= R - \fft{1}{4}\, F_\2^2 - (D-1)(D-2)\,
g^2\,.\label{emlagds}
\ee
Note that the kinetic term for $F_\2$ has the standard sign.  Using
the above analytical continuation, a cosmological solution can be
obtained from a superpotential construction, giving\footnote{A general
class of cosmological solutions in Einstein-Maxwell gravity with
a cosmological constant were also constructed in \cite{private}.}
\be ds_D^2=-H^{-1}\, dt^2 + H\, dx^2 + t^2\,
ds_{S^{D-2}}^2\,\,,\qquad H=g^2\, t^2 - (1 -
\fft{\ell^{D-3}}{t^{D-3}})^2\,. \ee
The solution is regular everywhere if the constraint
\be
g^2\,\ell^2 = (D-3)^2 \, (D-2)^{-\ft{2(D-2)}{D-3}}\,
\ee
is satisfied.  This cosmological solution interpolates between
dS$_2\times S^{D-2}$ in the infinite past to a dS$_D$-type spacetime
in the  infinite future, {\it i.e.}
\bea
\tau\rightarrow -\infty:&&
ds_D^2 = -d\tau^2 + e^{2\sqrt{D-2}\,g\,\tau}\, dx^2 + (D-2)^{\ft2{D-3}}\,
\ell^2\, d\Omega_{D-2}^2\,,\nn\\
\tau\rightarrow \infty:&&
ds_D^2 = -d\tau^2 + e^{2g\,\tau}\, (d\Omega_{D-2}^2 + dx^2)\,.
\eea
Similar results have been found in \cite{private}.  Here we only
present the cosmological solution explicitly for the Einstein-Maxwell
gravity with positive cosmological constant.  The cosmological
solutions for the more complicated system with scalar potentials can
be easily obtained, as we have demonstrated above, from AdS black
holes with analytical continuation.  The behavior of these solutions
are qualitatively the same as the one above.

Alternatively, we can consider the AdS black hole solution
(\ref{adsbh}). We consider the case $\epsilon=-1$ so that
$d\Omega_{D-2}^2$ is the metric of $H^{D-2}$, and perform the
following analytical continuation:
\be r\rightarrow i\,r\,,\qquad t\rightarrow x\,,\qquad
d\Omega_{D-2}^2=-ds_{{\rm dS}_{D-2}}^2\,. \ee
The resulting metric is given by
\be ds_D^2 = H\, dx^2+H^{-1}\, dr^2+r^2\, ds_{{\rm dS}_{D-2}}^2\,,
\label{bubble} \ee
where we have redefined
\be H=g^2\, r^2+1-\fft{M}{r^2}-\fft{Q^2}{r^4}\,. \ee
This is a non-BPS solution of AdS Einstein-Maxwell theory which
interpolates from $dS_{D-2}\times H^2$ at the near-horizon to an
$AdS_D$-type geometry with the boundary of $dS_{D-2}\times S^1$.
If $x$ is periodic then $r \ge r_H$ and the geometry is completely
smooth, where $r_H$ is the horizon radius of the AdS black hole
given by (\ref{adsbh}). Similar solutions have been discussed in
\cite{silver,bala,cvetic}.

\section{Conclusions}

      We have discussed how electric black holes in gauged
supergravity can be analytically continued to become magnetic
flux-branes.  We showed explicitly that these
flux-branes can be supersymmetric and regular everywhere,
interpolating between AdS$_{D-2}\times H^2$ at small distance to an
AdS$_D$-type geometry, with an AdS$_{D-2}\times S^1$ boundary,
in the asymptotic region. This differs from a
previously-known interpolation from AdS$_{D-2}\times H^2$ to
an AdS$_D$-type spacetime, where the asymptotic AdS$_D$-type geometry
has an M$_{D-3}\times H^2$ boundary.  Treating the AdS$_{D-2}$ as
the ``internal'' space, the new solution can be
viewed as a supergravity dual describing the renormalization group
flow of one-dimensional Euclidean conformal quantum mechanics.
Alternatively, the solution can be viewed as a smooth embedding of
AdS$_{D-2}$, suggesting a duality of certain conformal field
theories in different dimensions and signature.

          We have also used analytical continuation to obtain a
smooth cosmological solution of de Sitter Einstein-Maxwell
gravity, in which the kinetic term for the Maxwell field has the
standard sign. Although there is no supersymmetry, the solution
can arise from a first-order system.  It is regular at all times,
smoothly running from dS$_2\times S^2$ in the infinite past to a
dS$_D$-type geometry, with a Euclidean-signatured
$S^{D-2}\times S^1$ boundary, in the infinite future.

\section*{ACKNOWLEDGMENT}

        J.F.V.P. thanks the University of Kentucky and the University
of Cincinnati for hospitality during the intermediate phase of this
work.  C.N.P. thanks the Cambridge Relativity and Gravitation group,
and the Feza G\"ursey Institute in Istanbul, for hospitality during
the completion of this work.

\appendix
\section{General equations of motion}

      Let us consider a theory with the Lagrangian
\be
e^{-1}{\cal L}_D = R - \ft12 (\del\vec\phi)^2 - \ft14
\sum_{i}e^{\vec a_i\cdot \vec \phi}\, (F_\2^i)^2 - V(\vec \phi)
\,.
\ee
The equations of motion are given by
\bea
\square \vec\phi &=& \fft{\del V}{\del\vec\phi} +
\ft14 \sum_i\vec a_i\, (F_\2^i)^2\,
e^{\vec a_i\cdot \vec \phi}\,,\qquad d(e^{\vec a_i\cdot\vec\phi}\,
{*F_\2^i})=0\,,\nn\\
R_{\mu\nu} &=& \ft12 \del_\mu\phi\, \del_\nu\phi +
\ft{1}{D-2}\, V\, g_{\mu\nu} + \ft12 \sum_i e^{\vec a_i\cdot
\vec \phi}\, \Big((F^i)^2_{\mu\nu} - \ft1{2(D-2)} (F^i)^2\, g_{\mu\nu}
\Big)\,.
\eea

     We can construct a black hole solution with the following ansatz
\bea
ds_D^2 &=& -a^2 \, dt^2 + d\rho^2 + c^2\, d\Omega_{D-2}^2\,,\nn\\
e^{\vec a_i\cdot\vec\phi}\, {*F}_\2^i&=&\lambda_i\,
\Omega_{\sst{(D-2)}} \,,
\eea
where $d\Omega_{D-2}^2$ is the metric of a unit sphere $S^{D-2}$,
a unit hyperplane $H^{D-2}$ or a torus $T^{D-2}$.
Defining a vielbein basis $e^0=a\, dt$, $e^1=d\rho$ and
$e^\alpha=c\,\td e^\alpha$, we find that the spin connection
components are given by
\be
\omega_{01}=-\dot a\, dt\,,\qquad \omega_{1\alpha}=
-\dot c\, \td e^\alpha\,,\qquad
\omega_{\alpha\beta} = \td \omega_{\alpha\beta}\,,
\ee
and the Ricci tensor components are
\bea
&&R_{00} = \fft{\ddot a}{a} + (D-2)\, \fft{\dot a\,\dot c}{a\,c}\,,\qquad
R_{11}=-\fft{\ddot a}{a} - (D-2)\, \fft{\ddot c}{c}\,,\nn\\
&&R_{\alpha\beta}=\Big(-\fft{\dot a\,\dot c}{a\,c} -\fft{\ddot
c}{c} - (D-3)\, \fft{\dot c^2}{c^2} + \fft{(D-3)\,\ep}{c^2}\Big)
\delta_{\alpha\beta}\,, \eea
where $\epsilon=-1$, $1$ and $0$ for $H^{D-2}$, $S^{D-2}$ and
$T^{D-2}$, respectively. In the above, a dot denotes a derivative
with respect to $\rho$.

      Thus, the equations of the motion for the black hole are given by
\bea
&&\ddot {\vec\phi} + (\fft{\dot a}{a} + (D-2)\, \fft{\dot c}{c})\, \dot
{\vec\phi} =\fft{\del V}{\del\vec\phi}
- \ft12 c^{-2(D-2)}\, \sum_i\lambda_i^2
\,\vec a_i\, e^{-\vec a_i\cdot \vec\phi}\,,\nn\\
&&\fft{\ddot a}{a} + (D-2)\,\fft{\dot a\, \dot c}{a\,c} =
-\fft{V}{D-2} + \fft{(D-3)}{2(D-2)}\,
c^{-2(D-2)}\sum_i \lambda_i^2\,
e^{-\vec a_i\cdot \vec\phi}\,,\nn\\
&&-\fft{\ddot a}{a} - (D-2)\,\fft{\ddot c}{c} =
\fft{V}{D-2} + \ft12(\dot{\vec\phi})^2 -
\fft{(D-3)}{2(D-2)}\,
c^{-2(D-2)}\sum_i \lambda_i^2\,
e^{-\vec a_i\cdot \vec\phi}\,,\\
&&-\fft{\dot a\, \dot c}{a\,c} - \fft{\ddot c}{c} +
\fft{(D-3)\,\epsilon}{c^2} - \fft{(D-3)\,\dot c^2}{c^2} =
\fft{V}{D-2} + \fft{1}{2(D-2)}\, c^{-2(D-2)}\sum_i
\lambda_i^2\, e^{-\vec a_i\cdot \vec\phi}\,.\nn \label{eom} \eea

\end{document}